\documentclass[aps,showpacs,preprint,preprintnumber,amsmath,amssymb,ascmac,bm,12pt]{revtex4}
\usepackage{bm}
\usepackage[dvips]{graphicx}

\begin{document}

\title{
Cosmological Black Holes on Taub-NUT space in Five-Dimensional
Einstein-Maxwell Theory
}
\author{${}^{1}$Daisuke Ida\footnote{E-mail:daisuke.ida@gakushuin.ac.jp},
        ${}^{2}$Hideki Ishihara\footnote{E-mail:ishihara@sci.osaka-cu.ac.jp}, 
        ${}^{2}$Masashi Kimura\footnote{E-mail:mkimura@sci.osaka-cu.ac.jp},\\ 
        ${}^{2}$Ken Matsuno\footnote{E-mail:matsuno@sci.osaka-cu.ac.jp},
        ${}^{3}$Yoshiyuki Morisawa\footnote{E-mail:morisawa@keiho-u.ac.jp}
and
        ${}^{2}$Shinya Tomizawa\footnote{E-mail:tomizawa@sci.osaka-cu.ac.jp}}

\affiliation{
${}^1$ Department of Physics, Gakusyuin University, Tokyo 171-8588, Japan,\\
${}^2$ Department of Mathematics and Physics, Graduate School of Science, Osaka City University, 3-3-138 Sugimoto, Sumiyoshi, Osaka  558-8585, Japan\\
${}^3$ Faculty of Liberal Arts and Sciences, Osaka University of Economics and Law, Yao City, Osaka 581-8511, Japan.
}

\vspace{3cm}


\begin{abstract}
The cosmological black hole solution on the Gibbons-Hawking space has been constructed. 
We also investigate the properties of this solution in the case of a single black hole. 
Unlike the Kastor-Traschen solution, which becomes static 
solution in a single black hole, this solution is not static
even in a single black hole case.
\end{abstract}

\preprint{OCU-PHYS 259}
\preprint{AP-GR 38}
\pacs{04.50.+h, 04.70.Bw}

\maketitle

\section{Introduction}
The Majumdar-Papapetrou solution~\cite{MP} in Einstein-Maxwell theory 
describes an arbitrary number of extremely charged black holes in static equilibrium. The corresponding solution in Einstein-Maxwell theory with the positive cosmological term
was found by Kastor and Traschen~\cite{KT}.
The Kastor-Traschen solution describes the multi-black hole configuration 
in the de Sitter background. In particular, it realizes the analytic description of the
black hole collisions.
This implies that the Kastor-Traschen solution in general 
describes dynamical space-times in the sense that
there are no time-like Killing vector field. Nevertheless, it becomes static 
in the single black hole case, where it coincides with the extreme Reissner-Nordstr\"om-de Sitter space. The higher-dimensional generalization of the Kastor-Traschen solution is also
known~\cite{London}.

Recently, Kaluza-Klein black hole solution with a squashed event horizon in five-dimensional Einstein-Maxwell theory
was found by two of the present
authors~\cite{IM}. 
Thermodynamical properties of the Kaluza-Klein black holes are studied in \cite{CCO}, and
some generalizations are discussed in \cite{TWang,Yaz,BR}. The Kaluza-Klein black holes also admits multi-black hole configuration 
when black holes are extremely charged~\cite{IKMT}.
Though this Kaluza-Klein multi-black hole space-time
belongs to the five-dimensional Majumdar-Papapetrou class,
the four-dimensional base space becomes the Gibbons-Hawking multi-instanton space
rather than the usual flat space.
The spatial cross section of each black hole can be diffeomorphic to
the various lens spaces $L(n;1)$ ($n$: natural number) in addition to ${\rm S}^3$. In addition, a pair of static black holes on the Eguchi-Hanson space can be also considered \cite{IKMT2}.

The purpose of this article is to consider the effect of the 
cosmological constant
on the multi-black hole on the Gibbons-Hawking base space 
in five-dimensional Einstein-Maxwell theory and to investigate the global structure of this cosmological space-time.

The solution in consideration
is reduced 
to the Kaluza-Klein multi-black hole solution~\cite{IKMT}
by taking the limit of  zero cosmological constant, which is contained within supersymmetric solutions classified by Gauntlett et.al.~\cite{Ga}. It is found that, unlike the
Reissner-Nordstr\"om black hole in de Sitter background,
the space-time
becomes dynamical 
even in the single black hole case.

This article is organized as follows; 
In Sec.\ref{sec:solution}, we give a multi-black hole solution 
on the Gibbons-Hawking multi-instanton space
in the five-dimensional Einstein-Maxwell theory with a positive cosmological constant. 
In Sec.\ref{sec:pro}, we investigate the properties of this solution in the case of a single black hole. 
First, we show that there is no timelike Killing vector field. 
Next, we study the global structure of this solution. 
We give an example of the black hole space-time.

\section{Solutions }\label{sec:solution}

At first, we give a cosmological solution on the Gibbons-Hawking multi-instanton space. 
We consider five dimensional Einstein Maxwell system with a positive cosmological constant,
which is described by the action
\begin{equation}
S=\frac{1}{16\pi G_5}\int dx^5 \sqrt{-g} (R -4\Lambda-F_{\mu\nu}F^{\mu \nu} ),
\end{equation}
where $R$ is the five dimensional scalar curvature, $F_{\mu\nu}$ is the five-dimensional Maxwell field strength tensor, $\Lambda$ is the positive cosmological constant and $G_5$ is the five-dimensional Newton constant.

{}From this action, we write down the Einstein equation
\begin{equation}
R_{\mu\nu}-\frac{1}{2}Rg_{\mu\nu} +2g_{\mu \nu}\Lambda 
=
2\biggl(F_{\mu\lambda}{F_{\nu}}^{\lambda}-\frac{1}{4}g_{\mu\nu}F_{\alpha \beta}F^{\alpha\beta}\biggl),
\label{einsteineq}
\end{equation}
and the Maxwell equation
\begin{equation}
{F^{\mu \nu}}_{;\nu}=0.
\label{maxwelleq}
\end{equation}
Eqs.(\ref{einsteineq}) and (\ref{maxwelleq}) admit the solution whose metric and gauge potential one-form are
\begin{eqnarray}
ds^2&=&-H^{-2}dt^2 + H e^{\lambda t} ds_{\rm GH}^2, \\ \label{eq:solution}
{\bm A}&=&\pm \frac{\sqrt{3}}{2}H^{-1}dt,
\end{eqnarray}
with
\begin{eqnarray}
&&
H=1+\sum_{i}\frac{M_i}{e^{\lambda t}|\bm{x}-\bm{x}_{i}|},
\label{eq:harmonic}
\\
&&
d{s}^2_{\rm GH}
=V^{-1} ( d{\bm x}\cdot d{\bm x})
+V(d\zeta + {\bm\omega} )^2 ,
\\
&&
V^{-1} =\epsilon + \sum_{i}\frac{N_i}{|\bm{x}-\bm{x}_{i}|},
\\
&&
{\bm \omega}
=\sum_{i} N_i 
\frac{(z-z_{i})}{|\bm{x}-\bm{x}_{i}|}~
\frac{(x-x_i)dy -(y-y_i)dx}{(x-x_i)^2+(y-y_i)^2},
\end{eqnarray}
where $d{s}^2_{\rm GH}$ denotes the four-dimensional Euclidean Gibbons-Hawking space~\cite{GH},
$\bm{x}_i=(x_i,y_i,z_i)$ denotes position of the $i$-th 
NUT singularity with NUT charge $N_i$ 
in the three-dimensional Euclidean space, and $M_i$ is a constant. 
The constant $\lambda$ is given by $\lambda= \pm \sqrt{4\Lambda/3}$. 
The parameter $\epsilon$ is either $0$ or $1$. 
The base space is the multi-center Eguchi-Hanson space for $\epsilon=0$,
and the multi-Taub-NUT space for $\epsilon=1$. 
The $\epsilon=0$ solution
describes the coalescence of black holes
~\cite{IKT}.
Here, we study the $\epsilon=1$ solution and show
that it describes a black hole space-time.
 This solution is a generalization of five-dimensional Kaluza-Klein multi-black holes~\cite{IKMT}, and it is analogous to the Kaster-Traschen solution~\cite{KT,London}. 

\section{Properties of the single black hole}\label{sec:pro}
Let us consider the single black hole case.
Then, the metric takes the following simple form
\begin{eqnarray}
ds^2 &=& -\biggl(1 + \frac{M}{e^{\lambda t}R}\biggl)^{-2}dt^2 + \biggl(1 + \frac{M}{e^{\lambda t}R}\biggl)e^{\lambda t} ds^2_{\rm T-NUT},\label{eq:cosmological}
\\
ds^2_{\rm T-NUT} &=&\biggl(1+ \frac{N}{R}\biggl)(dR^2 + R^2d\theta^2 + R^2 \sin^2\theta d\phi^2 )
              + \biggl(1+ \frac{N}{R}\biggl)^{-1}N^2(d\psi + \cos \theta d\phi)^2 \nonumber 
\end{eqnarray}
where $ds^2_{\rm T-NUT} $ is the four-dimensional Euclidean self-dual Taub-NUT space.
The range of parameters are given by
 $0\le \theta \le\pi$, $0\le \phi \le2\pi$ and $0\le \psi \le 4\pi$. The Kastor-Traschen space-time which posseses the flat base space
becomes static in the single black hole case.
We show however that this property  does not hold in the present case.

\subsection{Nonsotatinarity}
The Kastor-Traschen solution in the single-black hole case is the
extreme Reissner-Nordstr\"om-de Sitter solution, which is static and spherically symmetric~\cite{BH,BHKT}. On the contrary, we show that the single black hole solution~(\ref{eq:solution})
is not stationary, but dynamical.
Let us seek for the stationary Killing vector field of 
the geometry (\ref{eq:cosmological}). 
For the present purpose, it is enough to exhaust the Killing vector fields for
the Gross-Perry-Sorkin (GPS) monopole solution with a cosmological constant
\begin{eqnarray}
ds^2 &=& -dt^2 + e^{\lambda t}ds^2_{\rm T-NUT},\label{eq:gps}
\end{eqnarray}
which is 
obtained by taking the limit $M\to 0$ of Eq.~(\ref{eq:cosmological}).
This is because the metric (\ref{eq:gps}) describes the far region ($R\to +\infty$)
from the black hole; if the original black hole has a stationary Killing vector field,
then the asymptotic form of the Killing vector would coincide with that of (\ref{eq:gps}).

All Killing vectors of the metric~(\ref{eq:gps}) are computed as follows, 
\begin{eqnarray}
&&\xi_1={\partial\over\partial \psi},\\
&&\xi_2={\partial\over\partial \phi},\\
&&\xi_3=\csc\theta\sin\phi{\partial\over\partial \psi}
+\cos\phi {\partial\over\partial \theta}
-\cot\theta\sin\phi {\partial\over\partial \phi},\\
&&\xi_4=\csc\theta\cos\phi {\partial\over\partial \psi}
-\sin\phi {\partial\over\partial \theta}
-\cot\theta \cos\phi {\partial\over\partial \phi}.
\end{eqnarray}
Thus, all the Killing vector fields are everywhere space-like.
Therefore, the single black hole solution (\ref{eq:cosmological}) is not stationary.

\subsection{GPS monopole with a cosmological constant}
\begin{figure}[htbp]
\begin{center}
\includegraphics[width=0.4\linewidth]{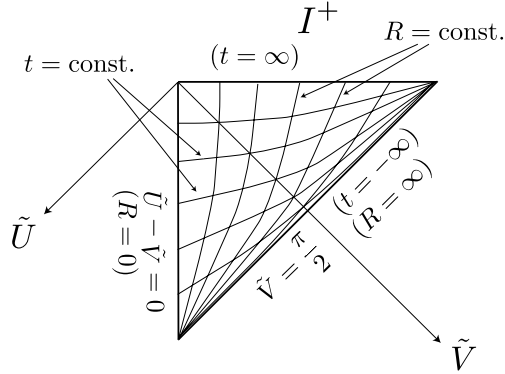}
\end{center}
\caption{The conformal diagram of the GPS monopole with a positive cosmological constant. The boundaries of this space-time consist of the timelike infinity $\tilde V=-\tilde U\ (0<\tilde V< \pi/2)$, $\tilde U=\tilde V\ (0<\tilde V<\pi/2)$ and the null surface $\tilde V=\pi/2,\ |\tilde U|<\pi/2$, which is not conformal infinity and cannot be extended analytically. }\label{fig:GPS}
\end{figure}

Here, we focus on
the case $\lambda >0$, which is called expanding chart. 
The contracting chart ($\lambda<0$) corresponds to its time-reversal ($t\to -t$).
Let us introduce the Regge-Wheeler tortoise type coordinate $R_*$ and a time coordinate $T$ are given by
\begin{eqnarray}
R_*=\sqrt{R(R+N)}+N\ln\biggl(\frac{\sqrt{R+N}+\sqrt{R}}{\sqrt{N}}\biggr),\quad T=\frac{2}{\lambda}e^{-\lambda t/2},\label{eq:R*}
\end{eqnarray}
where $R_*$ and $T$
range $0\le R_*\le \infty$ and $0\le T\le \infty$, respectively. 
Then, the metric (\ref{eq:gps}) takes the following form
\begin{eqnarray}
ds^2=\frac{4}{\lambda^2T^2}\Biggl[-dT^2+dR_*^2+\left(1+\frac{N}{R}\right)^{-1}R^2d\Omega_{\rm S^2}^2+\left(1+\frac{N}{R}\right)N^2(d\psi + \cos \theta d\phi)^2\Biggr].
\end{eqnarray}
Here, we define the double null coordinate $(u,v)$ such that
\begin{eqnarray}
v:=T+R_*,\quad u:=T-R_*,
\end{eqnarray}
where $v$ is the retarded outgoing null coordinate and $u$ is the retarded ingoing null coordinate.
Then, the metric can be written in the form
\begin{eqnarray}
ds^2=\frac{1}{\lambda^2(u+v)^2}\biggl[-dudv+R(R+N)d\Omega_{\rm S^2}^2+\frac{RN^2}{R+N}(d\psi + \cos \theta d\phi)^2\biggr].
\end{eqnarray}
To make coordinate ranges finite, define
\begin{eqnarray}
v=\tan \tilde V,\quad u=\tan \tilde U,
\end{eqnarray}
where $\tilde U$ and $\tilde V$ run the ranges of $-\pi/2\le \tilde U \le \pi/2$ and $-\pi/2\le \tilde V \le \pi/2$, respectively. The metric becomes
\begin{eqnarray}
ds^2&=&\frac{1}{\lambda^2\sin^2(\tilde U+\tilde V)}\biggl[-d\tilde Ud\tilde V+R(R+N)\cos^2\tilde U\cos^2\tilde Vd\Omega_{{\rm S}^2}^2\nonumber\\
   & &+\frac{RN^2}{R+N}\cos^2\tilde U\cos^2\tilde V(d\psi + \cos \theta d\phi)^2\biggr].\label{eq:GPS}
\end{eqnarray}
The $(R,t)$ chart covers the region ${\cal O}=\{(\tilde U,\tilde V)|\ \tilde U \le \tilde V$, $-\tilde U\le \tilde V,\tilde V\le \pi/2\}$. The conformal diagram is drawn in FIG.\ref{fig:GPS}. On the null surface $t \rightarrow -\infty, R \rightarrow \infty$, the ${\rm S}^1$ fiber seems to shrink to zero, the possibility of extension will be discussed in a future article.

\subsection{Single black hole solution}
\begin{figure}[htbp]
\begin{center}
\includegraphics[width=0.4\linewidth]{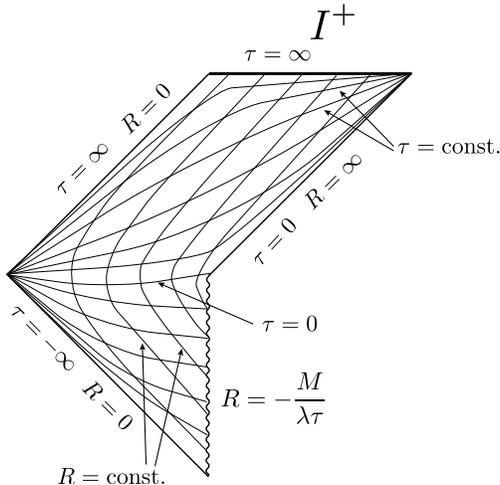}
\end{center}
\caption{The conformal diagram : The boundaries of this chart consist of (1) $I^+$:$\tau =\infty$ and $0<R<\infty$, (2) $\tau=\infty$ and $R=0$, (3) $\tau=-\infty$ and $R=0$, (4) $R=-M/(\lambda\tau)$, (5) $\tau=0$ and $R=\infty$. The spacelike hypersurface, $\tau =\infty$ and $0<R<\infty$, is the timelike infinity. $R=-M/(\lambda\tau)$ is a curvature singularity.
The null hypersurfaces, $\tau=\infty,\ R=0$ and $\tau=-\infty,\ R=0$, and the null surface $\tau =0$ and $R=\infty$ are not conformal boundary.}
\label{fig:BH}
\end{figure}

To study the $M>0$ case, introducing the coordinate $\tau:=\lambda^{-1}e^{\lambda t}$ ($\tau>0$), we rewrite the metric (\ref{eq:cosmological}) as
\begin{eqnarray}
ds^2&=&-\biggl(\lambda\tau+\frac{M}{R}\biggr)^{-2}d\tau^2\nonumber\\
   & &+\biggl(\lambda\tau+\frac{M}{R}\biggr)\biggl[\left(1+\frac{N}{R}\right)(dR^2+R^2d\Omega_{{\rm S}^2}^2)+\left(1+\frac{N}{R}\right)^{-1}N^2(d\psi+\cos\theta d\phi)^2\biggr].
\label{single}
\end{eqnarray}
This shows  that the range of $\tau$ can be extended to $-\infty<\tau<+\infty$.
The conformal diagram of this space-time is drawn in FIG.\ref{fig:BH}.
The sign choice of $\lambda$ corresponds to time reversal,
so that here we set $\lambda>0$, which is called expanding chart.
The boundaries of this chart consist of (1) $\tau =\infty$ and $0<R<\infty$, (2) $\tau=\infty$ and $R=0$, (3) $\tau=-\infty$ and $R=0$, (4) $R=-M/(\lambda\tau)$. (5) $\tau=0$ and $R=\infty$. The spacelike hypersurface denoted by $I^+$, $\tau =\infty,\ 0<R<\infty$ is the timelike infinity. The null hypersurface, $\tau=\infty$ and $R=0$, corresponds to the event horizon if we appropriately extend the space-time across this surface. 
The timelike curve $R=-M/(\lambda \tau)$ corresponds to a curvature singularity. 
The null surfaces $\tau=-\infty, ~R=0$ and $\tau=0,~R=\infty$ are the boundaries of the $(\tau,R)$ chart.

Here let us consider the global structure of this space-time.
As to static black hole solutions, the analytic extension across the boundaries of a chart can be done by
Edinton-Finkelstein coordinate~\cite{Wa}.
However, it is difficult to do such extension for our solutions since they are not static but dynamical as mentioned above, and one cannot solve the null geodesic equations easily. 
Now, we give one of extensions, which 
can be regarded as a black hole space-time, at least, within the restricted ranges of the parameters.

In the single black hole solution case, the form of the metric resembles that of the five-dimensional Reissner-Nordstr\"om-de Sitter
in the neighborhood of $R=0$, as mentioned below. For this reason, we apply the Edinton-Finkelstein-like coordinate to our solution to extend the space-time across $R=0$. The metric of five-dimensional Reissner-Nordstr\"om-de Sitter solution with $m=\sqrt3 |Q|/2$ 
can be written in the cosmological coordinate~\cite{London} as follows,
\begin{eqnarray}
ds^2 &=& -\bigg( \lambda  \tau + \frac{m}{r^2}\bigg)^{-2}d\tau^2 
+ \bigg( \lambda  \tau + \frac{m}{r^2}\bigg)
\bigg[
dr^2 
+ \frac{r^2}{4}d\Omega_{{\rm S}^2}^2 + \frac{r^2}{4}(d\psi + \cos \theta d\phi )^2
\bigg].\label{eq:RNdS}
\end{eqnarray}
To compare this metric with that of our solution, let us introduce the coordinate
\begin{eqnarray}
R:=\frac{r^2}{4 N},
\end{eqnarray}
where it should be noted that the parameter $N$ does not have a physical meaning.
In fact, it can be absorbed in the rescale of $R$.
The metric (\ref{eq:RNdS}) takes the form of
\begin{eqnarray}
ds^2 &=&
 -\left(\lambda \tau + \frac{M}{R}\right)^{-2}d\tau^2 
\notag
\\&&+ 
\left(\lambda \tau + \frac{M}{R}\right)
\left[
\left(\frac{N}{R}\right)(dR^2 + R^2d\Omega_{{\rm S}^2}^2)
+
\left(\frac{N}{R}\right)^{-1}N^2(d\psi + \cos \theta d\phi)^2
\right],\label{eq:RNdS2}
\end{eqnarray}
where we introduced $M:=m/(4N)$. 
We note that comparing with eqs (\ref{single}) and (\ref{eq:RNdS2}), the behavior of our solution in the neighborhood of $R=0$ is equal to that of the five dimensional Reissner-Nordstr\"om-de Sitter solution (\ref{eq:RNdS2}). 
In the case of the five-dimensional Reissner-Nordstr\"om-de Sitter solution (\ref{eq:RNdS2}), there exits
the coordinate across the horizon $(\tilde{R},v)$ as follows~\cite{London}
\begin{eqnarray}
 \tilde{R}^2 &:=& 4N\lambda \tau R + 4NM,\label{eq3}
\\
 v  &:=& \frac{\ln \lambda \tau}{\lambda}+ \int\frac{d\tilde{R}}{W} + \Delta(\tilde{R}), \label{eq32}
\end{eqnarray}
where the functions $\Delta(\tilde{R})$ and $W(\tilde{R})$ are defined by the equations
\begin{eqnarray}
\frac{ d\Delta(\tilde{R})} {d\tilde{R}}~  &&= -\frac{\lambda \tilde{R}^3}{2(\tilde{R}^2-m)}
\frac{1}{W(\tilde{R})},
\\
W(\tilde{R}) &&:= \left(1-\frac{m}{\tilde{R}^2}\right)^2 - \frac{\lambda^2 \tilde{R}^2}{4}
=\left(
1-\dfrac{m}{\tilde{R}^2}
+
      \dfrac{\lambda \tilde{R}}{2}
\right)
\left(
1-\dfrac{m}{\tilde{R}^2}
-
      \dfrac{\lambda \tilde{R}}{2} 
\right)
.
\end{eqnarray}
Then, in the coordinate $(\tilde R,v)$, the metric can be written in the form of
\begin{eqnarray}
ds^2
&=&
-Wdv^2
+2 dv d\tilde{R}
+\frac{\tilde{R}^2}{4}\left[d\Omega_{{\rm S}^2}^2 +(d\psi + \cos \theta d\phi)^2 \right].
\end{eqnarray}
In this coordinate $(\tilde R,v)$, three horizons correspond to the three positive roots of the equation $W=0$,
i.e. the cubic equations
\begin{eqnarray}
\pm\lambda \tilde R^3-2\tilde R^2-2m=0.\label{eq:cub}
\end{eqnarray}
This equation has three positive roots when the inequalities
\begin{equation}
0<m\lambda^2<{16\over 27}
\end{equation}
is satisfied.
Here, we denote these three roots as $\tilde R_I<\tilde R_H<\tilde R_C$, 
where $\tilde R=\tilde R_H$ correspond to the black hole horizon. 
Since the metric on the black hole horizon $\tilde R=\tilde R_H$ becomes
\begin{eqnarray}
ds^2 =
2 dv d\tilde{R}
+\frac{\tilde{R}_H^2}{4}\left[d\Omega_{{\rm S}^2}^2 +(d\psi + \cos \theta d\phi)^2 \right],
\end{eqnarray}
whose component takes the finite value there, 
then the space-time is extended to the region $\tilde R<\tilde R_H$ inside the black hole horizon.

\begin{figure}[htbp]
\begin{center}
   \includegraphics[width=0.5\linewidth]{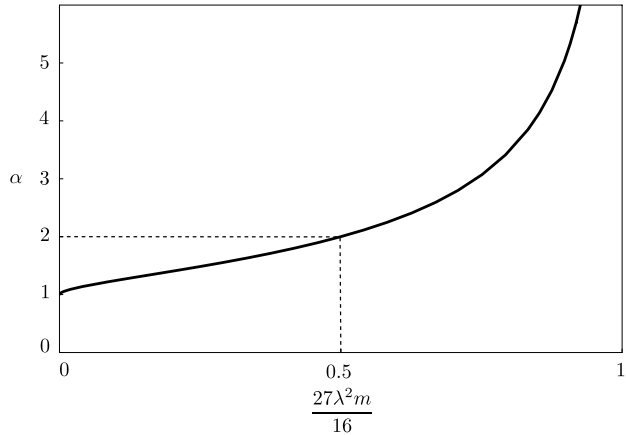}
\end{center}
\caption{
This graph shows how $\alpha$ depends on the parameter $\lambda^2 m$
for $0 \le \lambda^2m < 16/27$.}
\label{fig:n}
\end{figure}

\begin{figure}[htbp]
\begin{center}
   \includegraphics[width=0.5\linewidth]{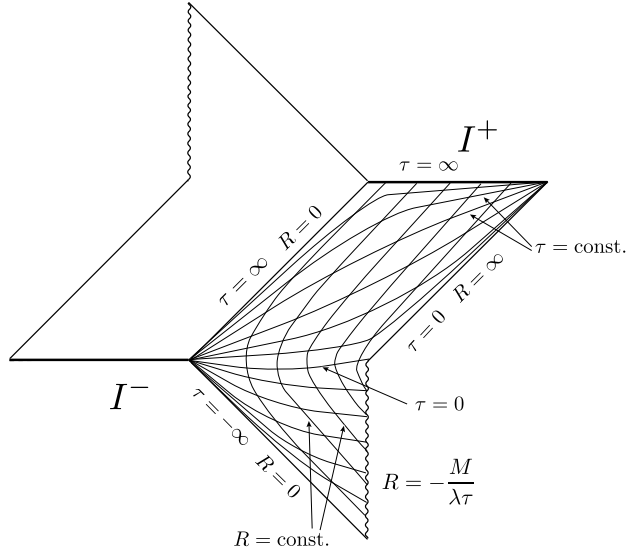}
\end{center}
\caption{Global structure of extended space-time.
The null hypersurface $R=0,\tau=\infty$ is the event horizon.
}
\label{fig:extension}
\end{figure}

Next, let us focus on the extension of the region covered with the chart $(\tau,R)$ in our solution (\ref{single})
, in particular, the extension across $R=0$ and $\tau=\infty$.  
To do so let us introduce the same coordinate as $(\tilde R,v)$ 
given by (\ref{eq3}) and (\ref{eq32}).
Then the metric of our solution (\ref{single}) takes the form
\begin{eqnarray}
ds^2
&=&
-Wdv^2
+2 dv d\tilde{R}
+\frac{\tilde{R}^2}{4}\left(
f
d\Omega_{{\rm S}^2}^2 +\frac{1}{f}(d\psi + \cos \theta d\phi)^2 \right)
\\&&+  (f-1)\frac{\lambda^2\tilde{R}^2}{4}\left( \frac{(1-m/\tilde{R}^2)+\lambda  \tilde{R}/2}{(\lambda  \tilde{R}/2) }\frac{1}{W}d\tilde{R}
-dv 
\right)^2,
 \notag
\end{eqnarray}
where $f$ is given by
\begin{eqnarray}
f:=1+\frac{R}{N}.
\end{eqnarray}
Here we note that $f\simeq 1$ near $R\simeq 0$. This coordinate seems to be spaned across the horizon, but actually it is a good coordinate only when $8/27\le m\lambda^2<16/27$ holds.
To see this we investigate the behavior of metric component near the horizon
which is located $\tilde{R}=\tilde R_H$, where $\tilde R_H$ is given by the root of Eq.(\ref{eq:cub}).
Then near the horizon, a function $W$ behaves as
\begin{eqnarray}
W  \sim (\tilde{R}-\tilde{R}_H),
\end{eqnarray}
and the behavior of the function $f-1$ is
\begin{eqnarray}
f-1= \frac{1}{4N^2}(\tilde{R}^2 -m)\exp{\left[-\lambda \left(v - \int\frac{d\tilde{R}}{W} + \Delta(\tilde{R})\right)
\right]}
\sim (\tilde{R}-\tilde{R}_H)^\alpha,
\end{eqnarray}
where the exponent $\alpha$ is given by
\begin{equation}
\alpha=\lim_{R \rightarrow R_H} \frac{\lambda}{
\left[
\left(1-\dfrac{m}{\tilde{R}^2}\right)-\dfrac{\lambda\tilde{R}}{2}
\right]
}
\frac{\tilde{R}^2}{(\tilde{R}^2-m)}(\tilde{R}-\tilde{R}_H).
\end{equation}
If the component $g_{\tilde R\tilde R}\sim(\tilde R-\tilde R_H)^{\alpha-2}$ does not diverge on the horizon $\tilde R=\tilde R_H$, it is also the case for the other components.
In fact, the other components behave as $O((\tilde R-\tilde R_H)^{\alpha-1})$ at most. Hence, it is sufficient to investigate the behavior of the component $g_{\tilde{R}\tilde{R}}$. 
The FIG.\ref{fig:n} shows how the value of $\alpha$ depends on the parameter $\lambda^2 m$.
{}From this graph, we see that $\alpha$ can take the value of $\alpha \ge 2$ for $8/27\le \lambda^2m < 16/27$, where the component $g_{\tilde R\tilde R}$ becomes finite at $\tilde R=\tilde R_H$.
Especially the metric is analytic at $\tilde R=\tilde R_H$ if $\alpha$ is a natural number grater than $1$.
FIG.~\ref{fig:extension} shows the conformal diagram of the extended space-time.We see that $\tilde R=\tilde R_H\ (R=0)$ corresponds to an event horizon, which is the boundary of the causal past of $I^+$, therefore our solution describes a black hole space-time.

\section{Summary and Discussion}\label{sec:final}
In this article, we have constructed the black hole solution on 
the self-dual Gibbons-Hawking space in the five-dimensional Einstein Maxwell theory with a positive cosmological constant. 
 In particular, we have studied the solution on the self-dual Taub-NUT base space in expanding phase and given an extension of space-time which describes a black hole space-time. 

We have also investigated the geometrical structure of this space-time
within the certain range of the parameters. 
In the neighborhood of the event horizon, the metric of our solution is equal to that of the five-dimensional Reissner-Nordstr\"om de-Sitter solution. Therefore, the spatial topology of a black hole horizon is ${\rm S}^3$. 
The asymptotic structure is described by the GPS monopole 
with a positive cosmological constant.
Remarkably, the solution has no timelike Killing vector in even a single black hole case and a cosmological GPS-monopole case,
unlike the Kastor-Traschen solution, which reduces to the static Reissner Nordstr\"om-de Sitter solution in a single black hole case.

Though it is expected that the boudaries of the chart $\tau=-\infty,R=0$ and $\tau=0,R=\infty$ should be extended, we do not have successful attempts in extending the spacetime across the surfaces. It is an open issue and may be interesting as a future work. In this article, we have focused on the single horizon solution.
While it is  expected that the multi-black hole solution
describe coalescing black holes. The grobal structure on higher-dimensional Kastor-Traschen solutions, which describe coalescing multi-black holes, is discussed in Ref.\cite{A}.
This case in our solution will be discussed in a future article.

\section*{Acknowledgments}
We are grateful to K. Nakao and H. Kodama for useful discussions. 
This work is supported by the Grant-in-Aid
for Scientific Research No.14540275 and No.13135208.

\end{document}